\documentclass[a4paper,structAbstract]{aa}  

\usepackage{natbib}
\bibpunct{(}{)}{;}{a}{}{,}
\usepackage{graphicx}
\usepackage{amsmath}	
\usepackage[usenames]{color}	
\usepackage{epstopdf}	
\usepackage{txfonts}
\usepackage{amssymb}
\usepackage{color}
\usepackage[justification=centering]{caption}
\usepackage[utf8]{inputenc}
\usepackage{geometry} 
\usepackage[parfill]{parskip} 
\usepackage{amssymb}
\usepackage{array}
\usepackage{url}
\usepackage{amsfonts}
\usepackage[colorlinks=true,linkcolor=black, urlcolor=black, citecolor=black]{hyperref}
\usepackage{hyperref} 
\usepackage{graphicx}
\usepackage{xcolor}
\usepackage{ulem}
\usepackage{epsfig}
\colorlet{rouge}{red!70!darkgray}

\begin{document}
\title{Constraints on the structure of $16$ Cyg $A$ and $16$ Cyg $B$ using inversion techniques}
\author{G. Buldgen\inst{1}\and D. R. Reese\inst{2}\and M. A. Dupret\inst{1}}
\institute{Institut d’Astrophysique et Géophysique de l’Université de Liège, Allée du 6 août 17, 4000 Liège, Belgium \and School of Physics and Astronomy, University of
Birmingham, Edgbaston, Birmingham, B15 2TT, UK.}
\date{November, 2014}
\abstract{Constraining additional mixing processes and chemical composition is a central problem in stellar physics as their impact on stellar age determinations leads to biases in our studies of stellar evolution, galactic history and exoplanetary systems. In two previous papers, we have shown how seismic inversion techniques could be used to offer strong constraints on such processes by pointing out weaknesses in current theoretical models. The theoretical approach having been tested, we now wish to apply our technique to observations. In that sense, the solar analogues 16CygA and 16CygB, being amongst the best targets in the Kepler field, are probably the current most well observed stars to test the diagnostic potential of seismic inversions.}
{We wish to use seismic indicators obtained with inversion techniques to constrain additional mixing processes in the structure of the components of the binary system 16Cyg. The combination of various seismic indicators will help to point out the weaknesses of stellar models and thus implies more constrained and accurate fundamendal parameters for these stars.}
{First, we will use the latest seismic, spectroscopic and interferometric observational constraints in the litterature for this system to determine suitable reference models independently for both stars. We will then carry out seismic inversions of the acoustic radius, the mean density and of a core conditions indicator. These additional constraints will be used to improve the reference models for both stars.}
{The combination of seismic, interferometric and spectroscopic constraints allows us to obtain accurate reference models for both stars. However, we note that a degeneracy exists for these models. Namely, changing the diffusion coefficient or the chemical composition within the observational values could lead to $5 \%$ changes in mass, $3 \%$ changes in radius and up to $8 \%$ changes in age. We used acoustic radius and mean density inversions to further improve our reference models then carried out inversions for a core conditions indicator, denoted $t_{u}$. Thanks to the sensitivity of this indicator to microscopic diffusion and chemical composition mismatches, we were able to reduce the mass dispersion to $2 \%$, namely $\left[ 0.96M_{\odot},1.0M_{\odot} \right]$, the radius dispersion to $1 \%$, namely $\left[ 1.188R_{\odot},1.200R_{\odot} \right]$ and the age dispersion to $3 \%$, namely $\left[7.0,7.4\right]$, for $16$CygA. For $16$CygB, $t_{u}$ offered a consistency check for the models but could not be used to reduce independently the age dispersion. Nonetheless, assuming consistency with the age of $16$CygA could help to further constrain its mass and radius. We thus find that the mass of $16$CygB should be between $0.93$ $M_{\odot}$ and $0.96$ $M_{\odot}$ and its radius between $1.08$ $R_{\odot}$ and $1.10$ $R_{\odot}$}{}
\maketitle

\section{Introduction}
In a series of previous papers (\citet{Buldgen} and (Buldgen et al. (submitted)), we have analysed the theoretical aspects of the use of seismic inversion techniques to characterize extra mixing in stellar interiors. Instead of trying to determine entire structural profiles, as was successfully done in helioseismology \citep{Basusound, Basudiff, BasuHelio} \footnote{Also see \citet{ReviewHelio} for an extensive review on helioseismology.}, we make use of multiple indicators, defined as integrated quantities which are sensitive to various effects in the structure. These indicators are ultimately new seismic constraints using all the available information provided by the pulsation frequencies. 
\\
\\
In this paper, we wish to apply our method to the binary system 16Cyg, observed by Kepler, for which data of unprecedented quality is available. Moreover, this system has already been extensively studied, particularly since the discovery of a red dwarf and a jovian planet in it \citep[see][]{Cochran}. Using Kepler data, this system has been further constrained by asteroseismic studies \citep{Metcalfe, Gruberbauer, Mathur}, interferometric radii have also been determined \citep[see][]{White} and more recently, Verma et al. have determined the surface helium abundance \citep{Verma} of both stars and \citet{Davies} have analysed their rotation profiles and tested gyrochronologic relations for this system.
\\
\\
The excellent quality of the Kepler data for these stars enables us to use our inversion technique to constrain their structure. We use the previous studies as a starting point and determine the stellar parameters using spectroscopic constraints from \citet{Ramirez} and \citet{Tucci}, the surface helium constraints from \citet{Verma} and the frequencies from the full length of the Kepler mission used in \citet{Davies} and check for consistency with the interferometric radius from \citet{White}. The determination of the stellar model parameters is described in Sect. \ref{secforward}. We carry out a first modelling process then determine the acoustic radius and the mean density using the SOLA technique \citep{Pijpers} adapted to the determination of these integrated quantities \citep[see][]{Buldgen,Reese}. In Sect. \ref{secinv}, we recall briefly the definition and purpose of the indicator $t_{u}$ and carry out inversions of this indicator for both stars.  We then discuss the accuracy of these results. Finally, in Sect. \ref{secconstraint}, we use the knowledge obtained from the inversion technique to provide additional and less model-dependent constraints on the chemical composition and microscopic diffusion in $16$CygA. These constraints on the chemical and atomic diffusion properties allow us to provide accurate, yet of course model-dependent, ages for this system, using the most recent observational data. The philosophy behind our study matches the so-called ``à la carte" asteroseismology of \citet{Lebreton} for HD$52265$, where one wishes to test the physics of the models and quantify the consequences of these changes. However, we add a substantial qualitative step by supplementing the classical seismic analysis with inversion techniques.

\section{Determination of the reference model parameters}\label{secforward}
\subsection{Initial fits and impact of diffusion processes}
\begin{figure}[t]
	\flushleft
		\includegraphics[width=9cm]{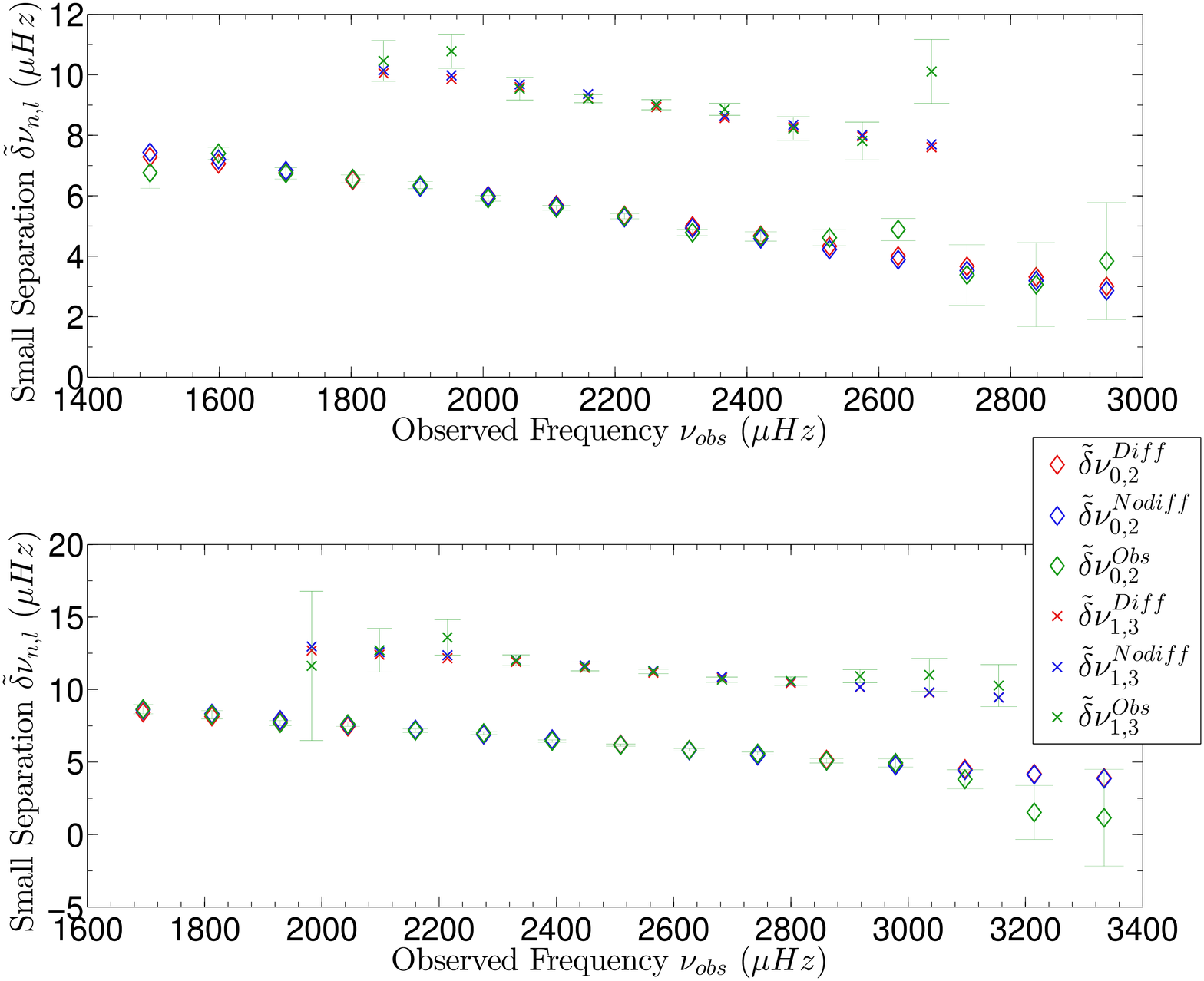}
	\caption{Upper panel: Fits of the small frequency separations $\tilde{\delta}_{02}$ and $\tilde{\delta}_{13}$ for $16$CygA. Lower panel: Same as the upper panel for $16$CygB. (Colour online) The observational values are the green symbols with error bars, the red symbols are associated with models including solar calibrated diffusion and the blue symbols are associated with models without diffusion.}
		\label{figFit}
\end{figure}

\begin{figure}[t]
	\flushleft
		\includegraphics[width=9cm]{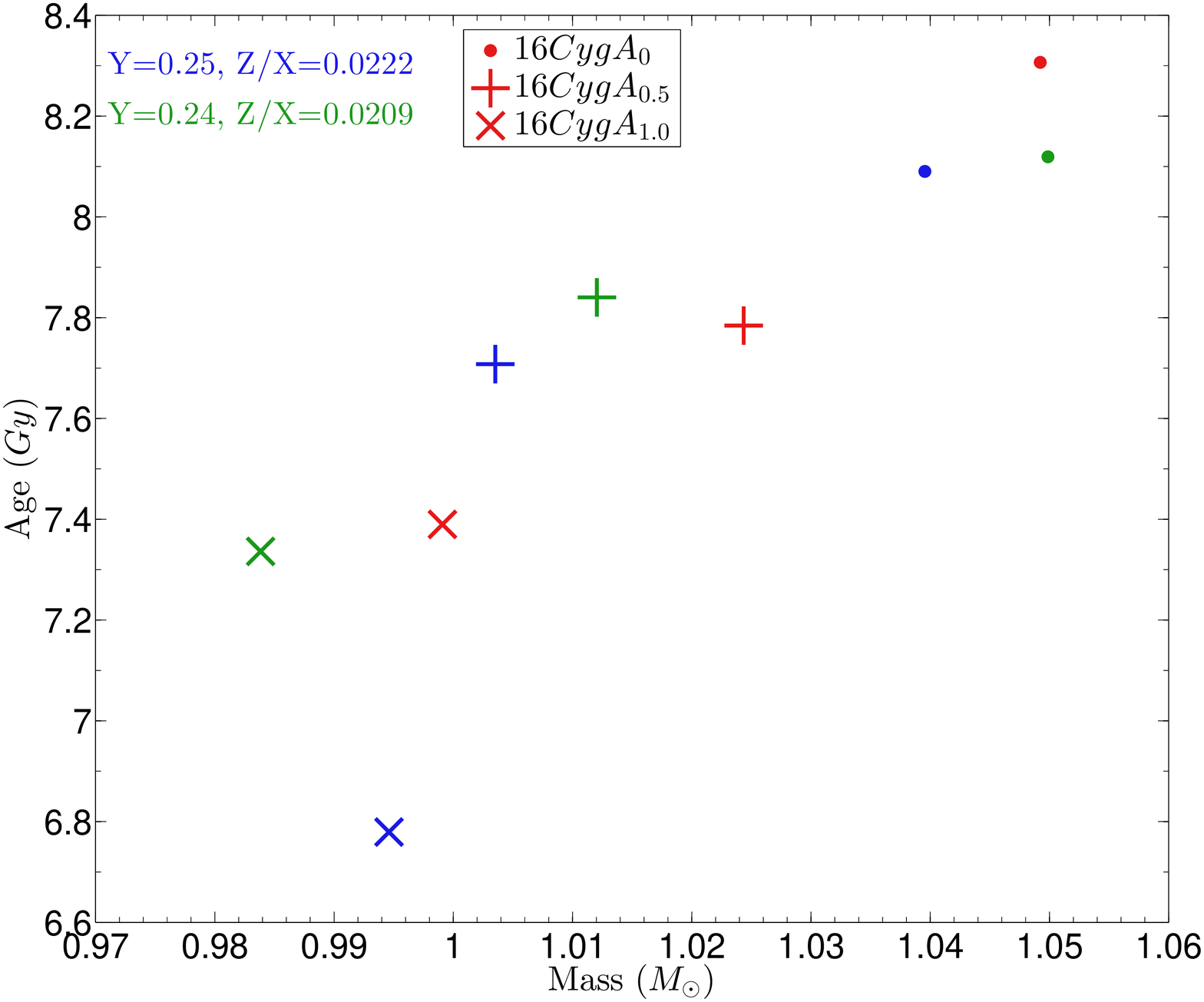}
	\caption{Effect of the progressive inclusion of diffusion in a model of $16$CygA. Each model still fits the observational constraints.}
		\label{figDiffAge}
\end{figure}

In this section, we will describe the optimization process that lead to the reference models for the inversions. We carried out an independent seismic modelling of both stars using the frequency spectrum from \citet{Davies}, which was based on $928$ days of Kepler data. A Levenberg-Marquardt algorithm was used to determine the optimal set of free parameters for our models. We used the Clés stellar evolution code and the Losc oscillation code \citep{ScuflaireCles, ScuflaireLosc} to build the models and calculate their oscillation frequencies. We used the CEFF equation of state \citep{CEFF}, the OPAL opacities from \citet{OPAL}, supplemented at low temperature by the opacities of \citet{Alexander} and the effects of conductivity from \citet{Potekhin} and \citet{Cassisi}. The nuclear reaction rates we used are those from the NACRE project \citep{Nacre} and convection was implemented using the classical, local mixing-length theory \citep{Bohm}. The empirical surface correction from \citet{Kjeldsen} was not used in this study. The cost function used to perform the minimization uses the well-known formula:
\begin{align}
\chi^{2}=\frac{1}{N-M}\sum_{i}^{N}\frac{\left(A^{i}_{obs}-A^{i}_{theo} \right)^{2}}{\sigma^{2}_{i}},
\end{align}
where $A^{i}_{obs}$ is an observational constraint, $A^{i}_{theo}$ the same quantity generated from the theoretical model, $\sigma_{i}$ is the observational error bar associated with the quantity $A^{i}_{obs}$, $N$ the number of observational constraints, and $M$ is the number of free parameters used to define the model. We can already comment on the use of the Levenberg-Marquardt algorithm, which is inherently a local minimization algorithm, strongly dependant on the initial values. In the following section, particular care was taken to mitigate the local character of the results since at least $35$ models were computed independently for each star, using various observational constraints and initial parameter values. As far as the error bars are concerned, we look at the dispersion of the results with changes in the physical ingredients rather than the errors given by the Levenberg-Marquardt algorithm.
\\
\\
We wish to emphasize that the use of other algorithms to select a reference model do not reduce the diagnostic potential of the inversions we will describe in the next sections. Indeed, inversions take a qualitative step beyond forward modelling techniques in the sense that they explore solutions outside of your initial model parameter space. 
\begin{figure}[t]
	\flushleft
		\includegraphics[width=9cm]{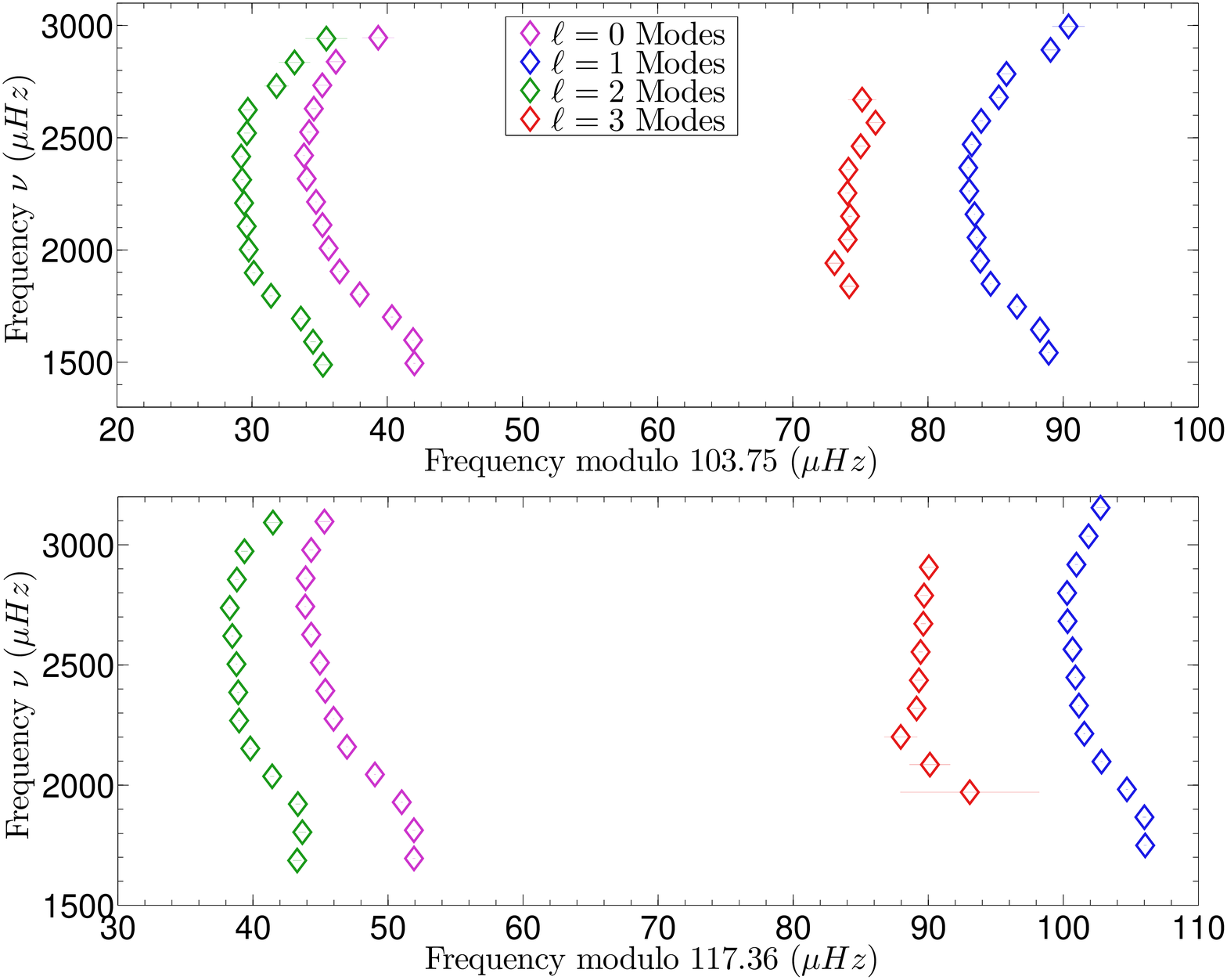}
	\caption{Upper panel: Observational echelle diagram of $16$CygA; Lower panel: Observational echelle diagram of $16$CygB; both plots illustrate the quality of the Kepler data for these stars. These are based on the frequencies obtained in \citet{Davies}. }
		\label{Echelle}
\end{figure} 
We used various seismic and non-seismic constraints in our selection process and we focussed our study on the importance of the chemical constraints for these stars. Indeed, there is a small discrepancy in the literature. In \citet{Verma}, a less model-dependent glitch fitting techniques was used to determine the surface helium mass fraction, $Y_{f}$. It was found to be between $0.23$ and $0.25$ for $16$CygA and between $0.218$ and $0.26$ for $16$CygB (implying an initial helium abundance, $Y_{0}$, between $0.28$ and $0.31$). In the seismic study of \citet{Metcalfe}, various evolutionary codes and optimization processes were used and the initial helium abundance was $0.25 \pm 0.01$ for a model including microscopic diffusion. In fact, the seismic study of \citet{Gruberbauer} already concluded that the helium mass fraction had to be higher than the values provided by \citet{Metcalfe}, which could result from the fact that they used $3$ months of Kepler data for their study. Therefore, the starting point of our analysis was to obtain a seismic model consistent with the surface helium constraint from \citet{Verma} and the metallicity constraint from \citet{Ramirez}. We started by searching for a model without including microscopic diffusion, and therefore the final surface abundances $Y_{f}$ and $Z_{f}$ are equal to the initial abundances $Y_{0}$ and $Z_{0}$. The metallicity can be determined using the following equation:
\begin{align}
&\left[\frac{Fe}{H} \right]= \log \left(\frac{Z}{X} \right)-\log \left(\frac{Z}{X} \right)_{\odot},
\end{align}
where $\left( \frac{Z}{X} \right)_{\odot}$ is the solar value consistent with the abundances used in the spectroscopic differential analysis.  We point out that in the spectroscopic study of \citet{Ramirez}, the ``solar'' references were the asteroids Cérès and Vesta. In this study, we used the $\left(\frac{Z}{X}\right)_{\odot}$ value from AGSS09 \citep{AGSS} to determine the value of the metallicity $Z$. From the error bars provided on these chemical constraints, we can determine a two dimensional box for the final surface chemical composition of the model (which is the initial chemical composition if the model does not include any extra mixing). A summary of the observed properties for both components is presented in table \ref{tabObs} as well as an observational echelle diagram in Fig \ref{Echelle} for both targets, illustrating the excellent quality of the seismic data for these stars.
\begin{table}[t]
\caption{Summary of observational properties of the system $16$CygA and $16$CygB considered for this study.}
\label{tabObs}
  \centering
\begin{tabular}{r | c | c  }
\hline \hline
 & \textbf{$16$CygA}& \textbf{$16$CygB} \\ \hline
\textit{R ($\mathrm{R_{\odot}}$)}& $1.22\pm 0.02$ & $1.12 \pm 0.02$ \\ 
\textit{$T_{\mathrm{eff,spec}}$ ($\mathrm{K}$)} & $5830 \pm 7$ & $5751 \pm 6$  \\ 
\textit{$T_{\mathrm{eff,phot}}$ ($\mathrm{K}$)} & $5839 \pm 42$ & $5809 \pm 39$  \\ 
\textit{$L_{\odot}$ $(\mathrm{L_{\odot}})$} & $1.56 \pm 0.05$ & $1.27 \pm 0.04$ \\
\textit{[Fe/H] (dex)} & $0.096$&$0.051$ \\
\textit{$Y_{f}$ (dex)} &$\left[0.23, 0.25\right]$&$\left[0.218, 0.260\right]$ \\
\textit{$<\Delta \nu>$ $(\mu Hz)$} &$103.78$&$117.36$ \\
\hline
\end{tabular}
\end{table}
An initial reference model without microscopic diffusion was obtained using the effective temperature, $T_{\mathrm{eff}}$, the arithmetic average of the large frequency separation $< \Delta \nu >$ and the individual small frequency separations $\delta \nu_{n,l}$. We did not include individual large frequency separations as these quantities are sensitive to surface effects in the frequencies and they would have dominated our cost-function. This would have been unfortunate since we want to focus our analysis on core regions. As we see from table \ref{tabfirstfitA}, this model was also able to fit constraints such as the interferometric radius from \citet{White} and the luminosity from \citet{Metcalfe} although these quantities were not included in the $\chi^{2}$ of the original fit. The agreement between the observed and theoretical seismic constraints is illustrated in Fig. \ref{figFit}. 
\begin{table}[t]
\caption{Optimal parameters obtained for $16$CygA.}
\label{tabfirstfitA}
  \centering
\begin{tabular}{r | c | c | c }
\hline \hline
 & \textbf{S$_{A,1}$}& \textbf{S$_{A,2}$}& \textbf{S$_{A,3}$} \\ \hline
 \textit{Mass ($\mathrm{M_{\odot}}$)}& $1.0523$ &$1.02494$& $1.00146$ \\
\textit{Radius ($\mathrm{R_{\odot}}$)}& $1.24$ & $1.22898$& $1.21838$ \\ 
\textit{Age ($\mathrm{Gyr}$)} &$8.23194$&$7.78380$ & $7.33473$ \\ 
\textit{$T_{\mathrm{eff}}$ ($\mathrm{K}$)} & $5825$ & $5802$ & $5801$ \\ 
\textit{$L_{\odot}$ $(\mathrm{L_{\odot}})$} & $1.58892$ & $1.53620$ & $1.50828$\\
\textit{$Z_{0}$} & $0.0165$&$0.019$ & $0.0205$ \\
\textit{$Y_{0}$} &$0.24$&$0.271$ & $0.2945$ \\
\textit{$\alpha_{\mathrm{MLT}}$} & $1.61757$ & $1.63975$ & $1.67229$ \\
\textit{$D$} & $0.0$ &$0.5$ & $1.0$ \\
\textit{$<\Delta \nu>$ $(\mu Hz)$} & $103.74$ &$103.79$ & $103.98$ \\
\textit{$\chi^{2}$} & $1.18$ &$1.19$ & $1.3$ \\
\hline
\end{tabular}
\end{table}
\begin{table}[t]
\caption{Optimal parameters obtained for $16$CygB.}
\label{tabfirstfitB}
  \centering
\begin{tabular}{r | c | c | c }
\hline \hline
 & \textbf{S$_{B,1}$}& \textbf{S$_{B,2}$}& \textbf{S$_{B,3}$} \\ \hline
 \textit{Mass ($\mathrm{M_{\odot}}$)}& $1.00839$ &$0.976542$& $0.942557$ \\
\textit{Radius ($\mathrm{R_{\odot}}$)}& $1.12292$ & $1.107$& $1.09847$ \\ 
\textit{Age ($\mathrm{Gyr}$)} &$8.16178$&$7.71671$ & $7.37336$ \\ 
\textit{$T_{\mathrm{eff}}$ ($\mathrm{K}$)} & $5749$ & $5742$ & $5739$ \\ 
\textit{$L_{\odot}$ $(\mathrm{L_{\odot}})$} & $1.23641$ & $1.1955$ & $1.17449$\\
\textit{$Z_{0}$} & $0.0151$&$0.0173$ & $0.0185$ \\
\textit{$Y_{0}$} &$0.24$&$0.273$ & $0.3015$ \\
\textit{$\alpha_{\mathrm{MLT}}$} & $1.567$ & $1.60264$ & $1.61543$ \\
\textit{$D$} & $0.0$ &$0.5$ & $1.0$ \\
\textit{$< \Delta \nu >$ $(\mu Hz)$} & $117.36$ &$118.00$ & $117.37$ \\
\textit{$\chi^{2}$} & $0.81$ &$0.85$ & $0.88$ \\
\hline
\end{tabular}
\end{table}
These results might seem correct, but since we did not even include microscopic diffusion, we should consider this model as rather unrealistic in terms of mixing processes\footnote{One should note that we do not imply here that microscopic diffusion is the only mixing process needed in a ``realistic model''.}. Therefore, we computed a few supplementary models assuming a final surface chemical composition of $Y_{f}=0.24$ and $\left(\frac{Z}{X}\right)_{f}=0.0222$ which included microscopical diffusion following the prescriptions of \citet{Thoul}. In this case, the fit was carried out using $5$ free parameters, the mass, the age, the mixing length parameter, denoted $\alpha_{\mathrm{MLT}}$, the initial hydrogen abundance, denoted $X_{0}$ and the initial metallicity, denoted $Z_{0}$. We used the same constraints as for the first fit without diffusion, supplemented by the constraints on the surface chemical composition, $Y_{f}$ and $(Z/X)_{f}$ providing direct and strong constraints on the initial chemical composition. 
\\
\\
The effect of diffusion was mainly to reduce the mass, age and radius of the model, as illustrated in Fig. \ref{figDiffAge}. This plot illustrates the effects of diffusion for various chemical compositions and diffusion velocities. The subscripts $0.0$, $1.0$ and $0.5$ are respectively related to a model without diffusion, with standard diffusion velocities and with half of these velocity values.  We will denote this factor $D$ in the tables presenting the results. Each color is associated with a particular surface chemical composition of these stars. All these models were fitted to the observed frequencies of their target. Therefore, the effect observed here is related to the impact of diffusion for a given model associated with a given set of frequencies. It is obvious that the reduction of the mass and radius are correlated since the mean density is kept nearly constant through the fit of the average large frequency separation. 
Therefore, the conclusion of this preliminary modelling process is that we obtain a degeneracy, meaning that we could build a whole family of acceptable models, inside the box of the chemical composition, with or without diffusion, that would be acceptable. This implies important uncertainties on the fundamental properties, as can be seen from the simple example in Fig. \ref{figDiffAge} for $16$CygA. In the following section, we will see how the use of inversion techniques and especially the inversion of $t_{u}$ can help us reduce this degeneracy and restrict our uncertainties on fundamental properties.
Even when considering diffusion based on the work of \citet{Thoul}, one should note that the diffusion velocities are said to be around $15-20 \%$ accurate for solar conditions. Therefore, in the particular case of $16$CygA, for which we have strong constraints on the chemical composition, one can still only say that the mass has to be between $0.97M_{\odot}$ and $1.07M_{\odot}$, that the radius has to be between $1.185R_{\odot}$ and $1.23R_{\odot}$ and that the age has to be between $6.8Gy$ and $8.3 Gy$ for this star. In other words, we have a $5 \%$ dispersion in mass, $3 \%$ in radius and $8 \%$ in age.

\subsection{Inversion of acoustic radii and mean densities}
\begin{figure*}[t]
	\flushleft
		\includegraphics[width=18cm]{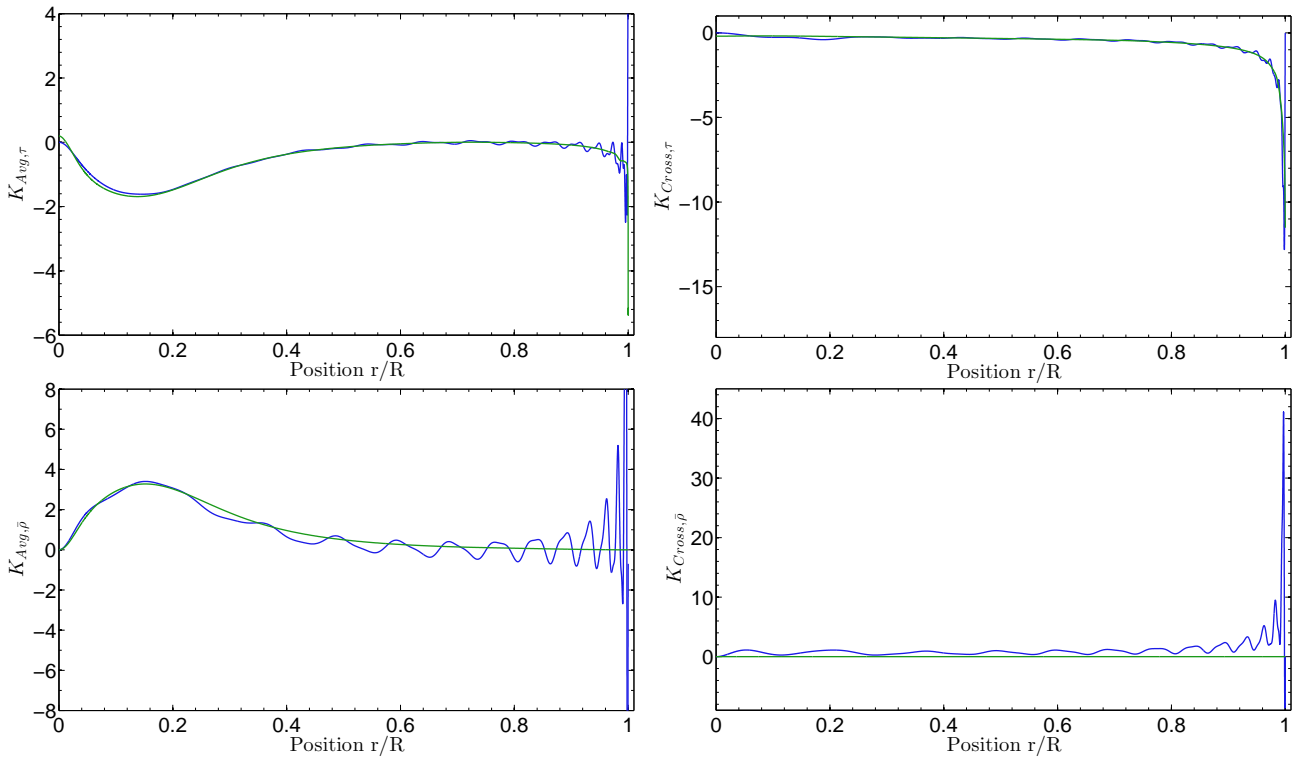}
	\caption{(Colour online) Upper panel: Example of Kernel fits for the inversion of the acoustic radius of $16$CygA (Averaging kernel on the left and cross-term kernel on the right). Lower panel: Kernel fits for the inversion of the mean density for $16$CygA (Averaging kernel on the left and cross-term kernel on the right). The target functions are in green and the SOLA kernels in blue.}
		\label{Kertaudens}
\end{figure*}

In this section, we will shortly present our results for the inversion of the mean density and the acoustic radius. First, we note that the inverted results for the mean density and the acoustic radius are slightly different from each other. There is a dispersion of around $0.5 \%$ for both $\bar{\rho}$ and $\tau$ depending on the reference model used for the inversion. We therefore consider that the results are respectively $\tau_{A}=4593 \pm 15 s$ and $\bar{\rho}_{A}=0.83 \pm 0.005 g/cm^{3}$ to be consistent with the dispersion we observe. For $16$CygB, we obtain similar results, namely $\tau_{B}=4066 \pm 15 s$ and $\bar{\rho}_{B}=1.066 \pm 0.005 g/cm^{3}$. The kernels are well fitted, as can be seen for a particular example in Fig. \ref{Kertaudens}.
\\
\\
This justifies the fact that linear inversions are said to be ``nearly model-independent". We emphasize that the physical ingredients for each model were different and that the dispersion of the results is smaller than $0.50\%$. Before the inversion, the dispersion of the mean density was of about $0.95\%$ and significantly different from the inversion results. In that sense, the model-dependency of these methods is rather small. However, the error bars determined by the simple amplification of the observational errors is much smaller than the model-dependency, thus one has to consider that the result is accurate within the dispersion due to the reference models rather than using the error bars given by the inversion. Nevertheless, this dispersion is small and therefore these determinations are extremely accurate. 
\\
\\
We also observed that including additional individual large frequency separations in the seismic constraints could improve the determination of both the acoustic radius and the mean density of the model. However, this can reduce the weight given to other seismic constraints and as we will see in the next section, we can improve the determination of reference models using directly the acoustic radius and the mean density as constraints in the fit. We also note that neither the mean density, nor the acoustic radius could help us disentangle the degeneracy observed in the previous section for the chemical composition and the effects of diffusion. Indeed, these quantities are more sensitive to changes in the mixing-length parameter, $\alpha_{MLT}$, or strong changes in metallicity. However, as will be described in the following section, they can be used alongside other inverted structural quantities to analyse the convective boundaries and upper layers of these stars.  
\subsection{Determination of new reference models}\label{secNewRef}
After having carried out a first set of inversions using the acoustic radius and the mean density, we carried out a supplementary step of model parameter determination, replacing the average large frequency separation by the acoustic radius and the mean density themselves. We obtained a new family of reference models which were slightly different from those obtained using the average large frequency separation. 
\\
\\
If we compare the model parameters obtained using $\tau$ and $\bar{\rho}$ for the model with $Y_{f}=0.24$ and $\left(Z/X \right)_{f}=0.0222$ with those obtained with $< \Delta \nu>$, presented in table \ref{tabfirstfitA}, we note that there is a tendency to slightly reduce the mass and to increase the mixing length parameter. What is more surprising is that when computing individual frequency differences between the observed stars and the reference models, we see that using the acoustic radius and the mean density allows us to obtain significantly better individual frequencies. This is a by-product of the use of inversion techniques that could be used to characterize stars in a pipeline such as what will be developed for the coming PLATO mission. 
\\
\\
Considering that these models are improved compared to what was obtained using the large frequency separation, we compute a family of models for different values of $Y_{f}$ and $\left( \frac{Z}{X} \right)_{f}$. For each particular chemical composition, we computed models with and without microscopic diffusion. The properties of some models of this family are summarized in table \ref{tabsecfitA}. As can be seen, some of the models do not reproduce well the results for the effective temperature or the interferometric radius. This means that we can use non-seismic constraint as indicators of inconsistent models in our study, although one should be careful on the conclusions derived from these quantities. For instance, the interferometric radii are different from the radii computed with the Clés models and some differences might result from the very definition of the radius. One should also note that these results are not totally incompatible since \citet{White} conclude that the radius of $16$CygA is $1.22 \pm 0.02 R_{\odot}$ and we find values around $1.185$ and $1.23$, outside the $1 \sigma$ errors for the lower part of our dispersion. The stellar luminosity is also dependent on these radii values and therefore should be considered with care. At the end of the day, the effective temperature can be constraining although there might be a slight difference stemming from discrepancies between the physical ingredients in the stellar atmosphere models used for the spectroscopic study of \citet{Ramirez} and \citet{Tucci} and those used in the Clés models in this paper. However, the inconsistencies observed for certain of these models are too important and therefore they should be rejected. The combination of all the information available will be described in Sect. \ref{secconstraint}. In the next section, we will use these models as references for our inversions of the $t_{u}$ indicator. One should note that this first step was beneficial since obtaining reference models as accurate as possible for these stars is the best way to obtain accurate results for the more difficult inversion of the $t_{u}$ indicator.
\section{Inversion results for the $t_{u}$ core conditions indivator}\label{secinv}
\subsection{Definition of the indicator and link to mixing processes}
\begin{figure*}[t]
	\centering
		\includegraphics[width=17cm]{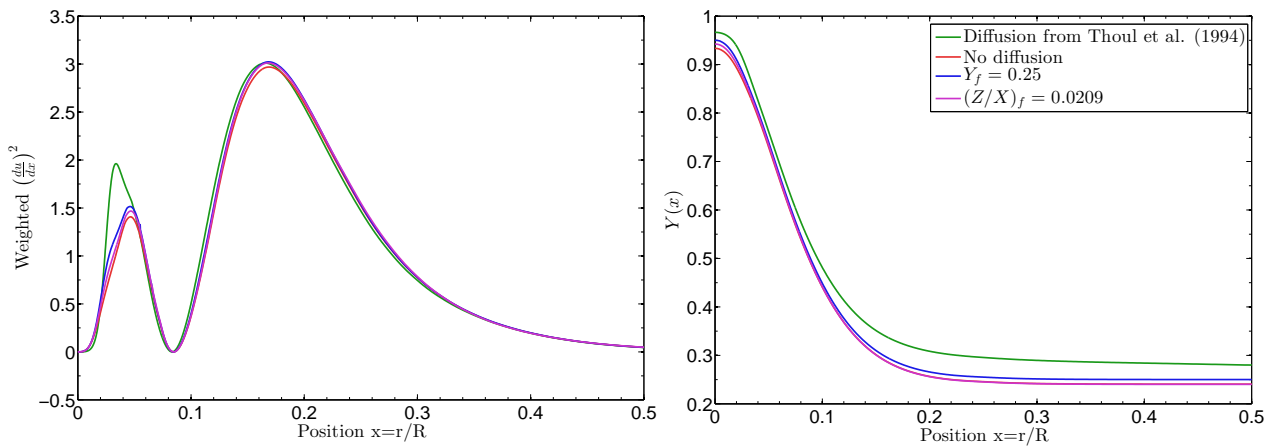}
	\caption{Left panel: Effect of diffusion, metallicity changes and helium abundance changes on the core regions for models \textbf{S$_{A,C1}$}, \textbf{S$_{A,C2}$}, \textbf{S$_{A,L1}$}, \textbf{S$_{A,U1}$} on the target function of $t_{u}$, since the quantity is integrated, the sensitivity is greatly improved. Right panel: the $Y(x)$ profile of these models is illustrated, thus showing the link between $t_{u}$ and chemical composition and thus, its diagnostic potential.}
		\label{Figdiff}
\end{figure*}

In Buldgen et al. (submitted), we defined and tested a new indicator for core conditions, applicable to a large number of stars\footnote{Provided that there is sufficient seismic information for the studied stars.} and very sensitive to microscopic diffusion or chemical composition mismatches in the core regions between the target and the reference model. The definition of this quantity was the following:
\begin{align}
t_{u}=\int_{0}^{R}f(r)\left(\frac{du}{dr} \right)^{2}dr \label{Eqtu},
\end{align}
where $u$ is the squared isothermal sound speed, defined as $u=\frac{P}{\rho}$ $f(r)$ is a weighting function defined as follows:
\begin{align}
f(r)=r \left( r-R \right)^{2} \exp \left(-7 \left(\frac{r}{R}\right)^{2}\right).
\end{align}
Due to the effects of the radius differences between the observed target and reference model, we noted that the quantity measured was $\frac{t_{u}}{R^{6}_{tar}}$, where $R_{tar}$ is the target radius. In Fig. \ref{Figdiff}, we illustrate the changes of the quantity due to the effects of diffusion for two of our reference models, having the same surface chemical composition and fitting the same observational constraints.
One can also see the effect of surface helium and metallicity changes on the  profile of the integrant of Eq. \ref{Eqtu}. The whole parameter set of these models is given in table \ref{tabsecfitA} as well as the explanation of the naming convention. The diagnostic potential of the $t_{u}$ inversion is therefore clear, although the weighting function could be adapted to suit other needs if necessary. The inversion of this integrated quantity can be made using both the $\left(u_{0},\Gamma_{1} \right)$ or the $\left(u_{0},Y \right)$ kernels.
\subsection{Inversion results for $16$CygA}\label{secresCygA}
\begin{figure*}[t]
	\flushleft
		\includegraphics[width=18cm]{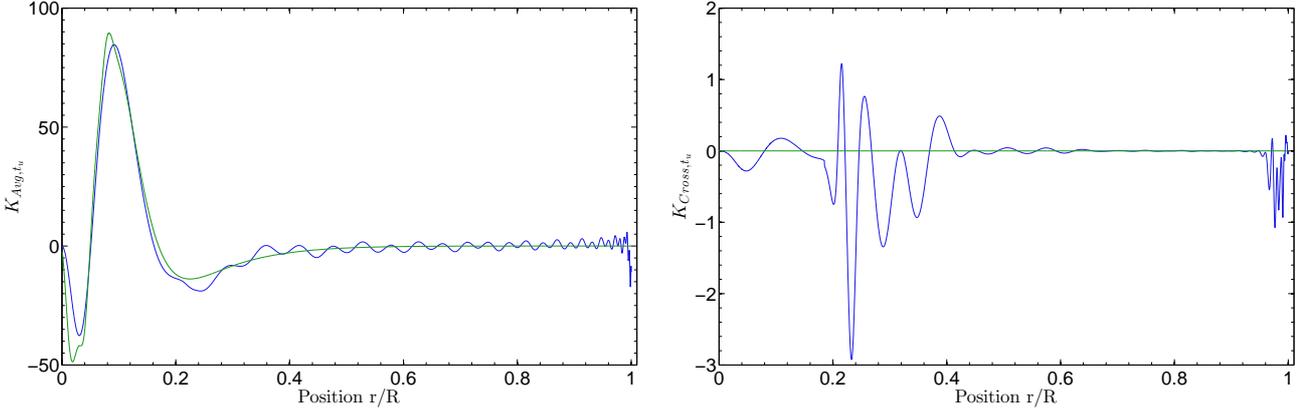}
	\caption{(Colour online) Example of kernel fits for the $t_{u}$ inversion. The left panel is associated with the averaging kernel and the right panel is associated with the cross-term kernel.  The target functions are in green and the SOLA kernels in blue.}
		\label{Kertu}
\end{figure*}

The inversion results are summarized in Fig. \ref{resultsbox} (they are represented as orange $\times$ in the $\bar{\rho}-\frac{t_{u}}{R^{6}}$ plot) and illustrated through an example of kernel fits in Fig. \ref{Kertu}. We tried using both the $\left(u_{0},\Gamma_{1} \right)$ and the $\left(u_{0},Y \right)$ kernels. The high amplitude of the $\Gamma_{1}$ cross-term leads us to present instead the results from the $\left(u_{0},Y \right)$ kernels although they are quite similar in terms of the inverted values. However, one should note that the error bars are quite important, and we have to be careful when interpreting the inversion results. 
\\
\\
This effect is due to both the very high amplitude of the inversion coefficients and the amplitude of the observational error bars. When compared to the somewhat underestimated error-bars of the acoustic radius and mean density inversion, it illustrates perfectly well why it is always said that two inversion problems can be completely different. In this particular case, using various reference models allows us to already see a trend in the inversion results. We clearly see that the value of $t_{u}$ for our reference models is too low and we see that the dispersion of the inversion results is rather low, despite the large error bars. One should also note that the quality of the kernel fit is also a great indicator of the quality of the inverted result. For most cases, the kernels were very well fitted and the low dispersion of the results means that there is indeed information to be extracted from the inversion. We will see how this behaviour is clearly different for $16$CygB. 
\\
\\
Nevertheless, one could argue that a small change of $t_{u}$ could be easily obtained through the use of diffusion, or chemical composition changes. We will see in Sect. \ref{secconstraint} how combining all the information along with new constraints from the inversion technique can be extremely restrictive in terms of chemical composition and diffusion processes. Indeed, $t_{u}$ should not be considered as a model-independent age determination or as an observed quantity disentangling all physical processes occurring in stellar cores. In fact, it is simply a nearly model-independent determination of a structural quantity optimized to more sensitive to any change in the physical conditions in stellar cores than classical seismic indicators. The amplitude of the error bars remind us that this sensitivity comes at a cost and in this study we will consider that having a reference model with a $\frac{t_{u}}{R_{ref}^{6}}\approx 3.2 $ or $3.3$ $\frac{g}{cm^{6}}$ will be acceptable if it still fits the other observational constraints.
\subsection{Inversion results for $16$CygB}\label{secrescygb}
\begin{figure}[t]
	\flushleft
		\includegraphics[width=9cm]{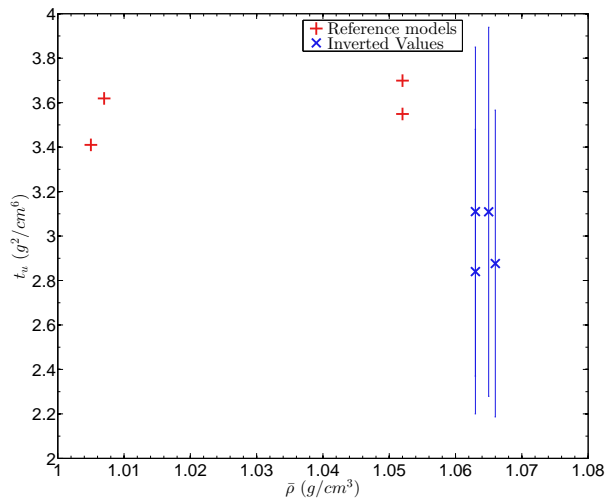}
	\caption{$t_{u}$ inversion results for $16$CygB. (Colour online) The red $+$ are the reference models and the blue $\times$ the inverted results. The lower $+$ are associated with the upper $\times$ and refer to models including solar-calibrated diffusion.}
		\label{resB}
\end{figure}

The case of $16$CygB is completely different. In fact, whereas the inversion for the acoustic radius and the mean density have been successful and we could build improved models for this star, the inversion of the $t_{u}$ indicator was less successful. The results were good, in the sense that the kernels are well fitted. However, we can see from Fig. \ref{resB} that the amplification of the observational error was too high to constrain the microscopic diffusion effects or the chemical composition. In fact, it is not surprising since the error bars on the observed frequencies are larger than for $16$CygA. 
\\
\\
As a matter of fact, the observational error dominate the inversion result, as can be easily shown in Fig. \ref{resB}. We see that the relative change in $t_{u}$ is smaller when microscopic diffusion is included in the model but this is due to the fact that the inversion result is closer to the reference value rather than the opposite. Thus, this means that $t_{u}$ can be used as a consistency check for future investigations, to ensure that we stay within the error bars of the inverted value, but we cannot hope to gain additional information for this star from this indicator. 
\section{Constraints on microscopic diffusion and chemical composition}\label{secconstraint}
\subsection{Reducing the age, mass and radius dispersion of $16$CygA}
\begin{figure*}[t]
	\centering
		\includegraphics[width=16cm]{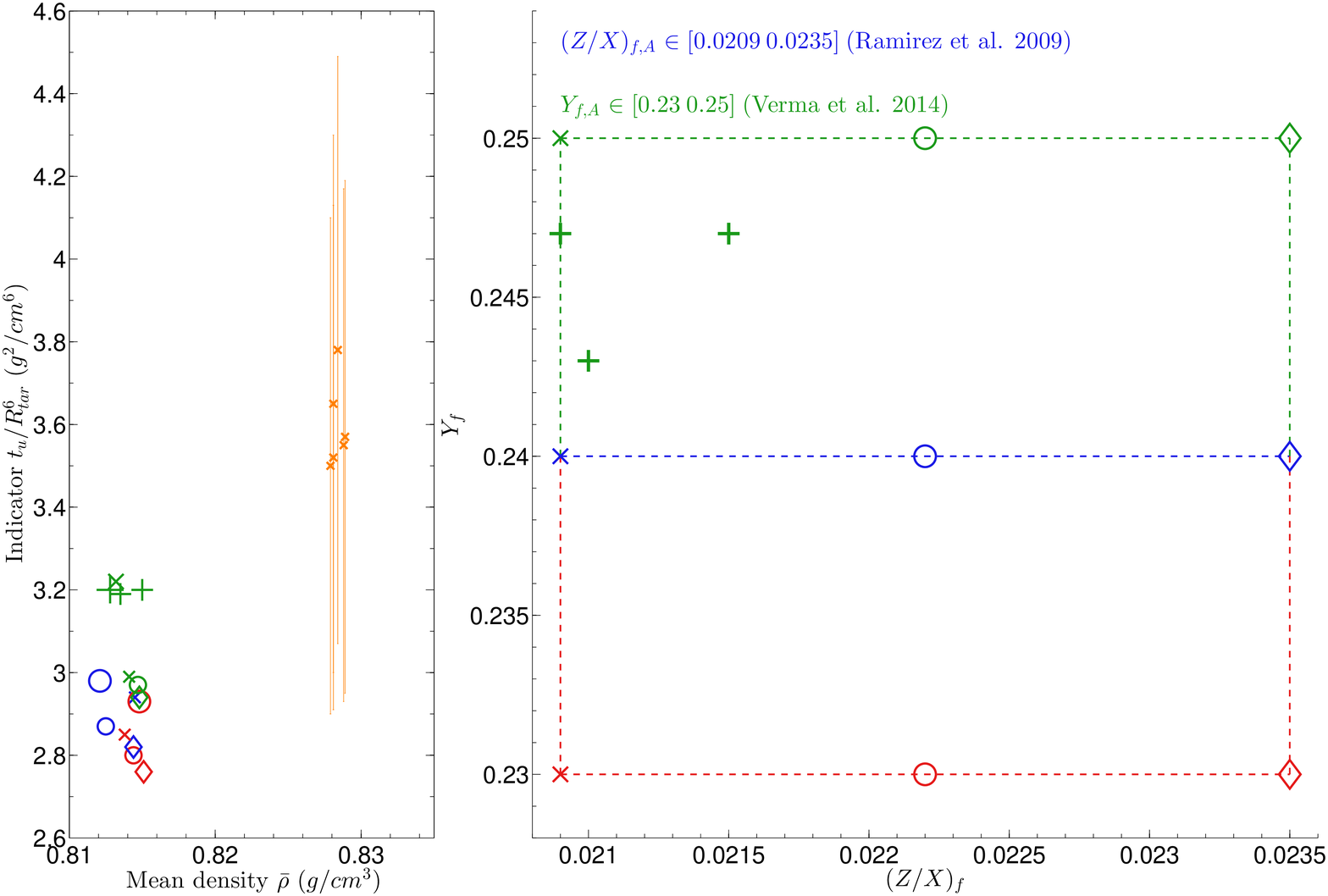}
	\caption{(Colour online) Results of the $t_{u}$ inversions for $16$CygA (left panel) and positions of the reference models in the chemical composition box derived from spectroscopic and seismic constraints (right panel). The orange $\times$ are the inversion results whereas the other symbols are associated with various reference models the positions of which are shown in the right-hand side plot. The colour is associated with the $Y_{f}$ value, the type of symbol with the $\left( \frac{Z}{X}\right)_{f}$ value, and the size of the symbol with the inclusion of diffusion.}
		\label{resultsbox}
\end{figure*}

In this section, we will use the information given by $t_{u}$ to further constrain chemical composition and microscopic diffusion. Previously, we always ensured that the reference models were inside the chemical composition box that was defined by the constraints on surface helium obtained by \citet{Verma} and the spectroscopic constraints obtained by \citet{Ramirez}. In Sect.\ref{secresCygA}, we concluded that our model should have at least a $\frac{t_{u}}{R_{ref}^{6}}\approx 3.2 $ or $3.3$ $\frac{g}{cm^{6}}$ or higher. The first question that arises is whether it is possible to obtain such values for $\frac{t_{u}}{R^{6}}$ given the constraints on chemical composition. The second question is related to the impact of microscopic diffusion. 
\\
\\
In fact, $t_{u}$ is a measure of the intensity of the $u_{0}$ gradients in the core regions. Thus, since $u_{0} \approx \frac{T}{\mu}$, where $T$ is the temperature and $\mu$ the mean molecular weight, including diffusion will increase the $\mu$ gradients, since it leads to the separation of heavy elements from lighter elements. It is then possible to increase significantly the diffusion speed of the chemical elements and to obtain a very high value of $t_{u}$ for nearly any chemical composition. However, in \citet{Thoul}, the diffusion speed are said to be $\approx 15-20 \%$ accurate, and suited for solar conditions. Moreover, since increasing diffusion also accelerates the evolution, we could also end up with models that are too evolved to fit simultaneously $t_{u}$, the chemical composition constraints and the seismic constraints. Looking at the parameters of our reference models, we note that we are indeed very close to solar conditions and we suppose that our diffusion speed should not be amplified or damped by more than $20 \%$. The results of this analysis are summarized in Fig. \ref{resultsbox}, which is a $\bar{\rho}-\frac{t_{u}}{R^{6}}$ plot where the reference models and the inverted results are represented. In what follows, we will describe more precisely our reasoning and refer to Fig. \ref{resultsbox} when necessary. We used a particular colour code and type of symbol to describe the changes we applied to our models. Firstly, colour is associated with the final surface helium mass fraction $Y_{f}$, namely, blue for $Y_{f}=0.24$, red if $Y_{f}<0.24$ and green if $Y_{f}>0.24$. Secondly, the symbol itself is related to the $\left(\frac{Z}{X}\right)_{f}$, namely, a $\times$ for $\left(\frac{Z}{X}\right)_{f} < 0.0222$, a $\circ$ for $\left(\frac{Z}{X}\right)_{f} = 0.0222$ and a $\diamond$ for $\left(\frac{Z}{X}\right)_{f} > 0.0222$. The size of the symbol is related to the inclusion of microscopic diffusion, for example the large blue and red dots in Fig. \ref{resultsbox} are related to models including microscopic diffusion. 
\\
\\
Since increasing diffusion should increase the $t_{u}$ value, we compute a model, with $Y_{f}=0.24$ and $\left( \frac{Z}{X} \right)_{f}=0.0222$, including diffusion from \citet{Thoul}, fitting the seismic constraints and the effective temperature. This model is represented by the large blue dot and we note that including diffusion improves the agreement, but is not sufficient to reach what we defined to be our acceptable values for $\frac{t_{u}}{R^{6}}$. This is illustrated by the fact that in Fig. \ref{resultsbox}, the large blue dot is above the small blue dot. Therefore we decide to analyse how $t_{u}$ depends on the chemical composition. To do so, we compute a model for each corner and each side of the chemical composition box. These models are represented in Fig. \ref{resultsbox} by the $\diamond$, $\circ$ and $\times$ of various colours. From these results, we see that increasing the helium content, namely considering that $Y_{f} \in \left[ 0.24, 0.25 \right]$ increases $t_{u}$, as does considering $\left(\frac{Z}{X}\right)_{f} \in \left[0.0209, 0.0222\right]$. In simpler terms, we see that the green dot and the blue cross are above the blue dot in Fig. \ref{resultsbox}.
The first tendency is quickly understood since increasing the helium abundance leads to higher central $\mu$ and therefore a local minimum in the $u_{0}$ profile. Because $t_{u}$ is based on $\left( \frac{du}{dx}\right)^{2}$, this does not imply a reduction of the value of the indicator, but an increase due to a secondary lobe developing exactly in the same way as what happens when including diffusion (see Fig. \ref{figDiffAge}). The second tendency can be understood by looking at the central hydrogen abundance. In this case, we see that the central hydrogen abundance is reduced and thus the mean molecular weight is increased and lead to a minimum in $u_{0}$ in the centre. One should note that this effect is not as intense as the change in helium but is still non-negligible. 
\\
\\
Therefore, our seismic analysis favours models that lie within $Y_{f} \in \left[ 0.24, 0.25 \right]$ and $\left( \frac{Z}{X}\right)_{f} \in \left[0.0209, 0.0222\right]$. Including diffusion in these models increases further the $\frac{t_{u}}{R^{6}}$ value and brings it in the range of the $3.2$, $3.3$ $\frac{g}{cm^{6}}$ values, which is much more consistent with the inversion results. These final models are represented in Fig. \ref{resultsbox} by the large green $+$. One should also note that an upper boundary can be drawn from the effective temperature and the seismic constraints. In other words, the fit of the other quantities can slightly increase the $\chi^{2}$ up to values of $1.6$ and thus slightly reduce the quality of the fit. This is not alarming but still means that one should not put all the weight of the fit of the model on the inversion results but try to find a compromise between seismic, spectroscopic, and inverted constraints. 
\begin{table}[t]
\centering
\caption{Accepted parameters obtained for $16$CygA when taking into account the constraints from the inversion of $t_{u}$.}
\label{tabcygAfinalfit}
\begin{tabular}{r | c}
\hline \hline
& \textbf{Accepted $16$CygA models} \\
\hline
$M$ $(\mathrm{M_{\odot}})$& $0.96-1.00$ \\
Age $(\mathrm{Gy})$& $7.0-7.4$\\
$Y_{0}$ $\mathrm{(dex)}$& $0.30-0.31$\\
$Z_{0}$ $\mathrm{(dex)}$& $0.0194-0.0199$\\
$D$ & $1.00-1.15$ \\
$\alpha_{\mathrm{MLT}}$ & $1.75-1.90$ \\
$L$ $(\mathrm{L_{\odot}})$ & $1.49-1.56$ \\
$R$ $(\mathrm{R_{\odot}})$& $1.19-1.20$ \\
\hline
\end{tabular}
\end{table}
Considering models that fulfil all these constraints, we are able to reduce the degeneracy previously observed. We thus conclude that the mass of $16$CygA must be between $0.97M_{\odot}$ and $1.0M_{\odot}$ and its age must be between $7.0Gy$ and $7.4Gy$. These values are subject to the hypotheses of this study and depend on the physics used in the stellar models (opacities, nuclear reaction rates, abundances). We recall here that there is no way to provide a seismic fully model-independent age, but inversions allow us at least to check the consistency of our models with less-model dependent structural quantities. These consistency checks can lead to a refinement of the model parameters and in this particular case to constraints on microscopic diffusion.
\\
\\
For the sake of completion, we also analysed the importance of the abundances used to build the model. In fact, the [Fe/H] constraint being extremely dependent on the solar $\left(\frac{Z}{X}\right)_{\odot}$, we wanted to ask the question of whether the inversion would have also provided a diagnostic if we had used the GN$93$ abundances to determine the metallicity. Using these abundances and the associated $\left(\frac{Z}{X}\right)_{\odot}$ which is equal to $0.0244$, one ends up with models having much higher metallicities, of the order of $0.0305$ when no diffusion is included in the model. In fact we ended up with the same tendencies in the chemical composition box, but with completely different values of $\left( \frac{Z}{X}\right)$, implying slightly higher masses of around $1.03M_{\odot}$ and slightly lower ages around $6.8Gy$. However, when carrying out the $t_{u}$ inversion, we noted that we still had to increase the helium content, include diffusion, and reduce the $\left(\frac{Z}{X}\right)$. The interesting point was that even the lowest $\left(\frac{Z}{X}\right)$, associated with the highest $Y_{f}$ with increased diffusion could not produce a sufficiently high value of $t_{u}$. In that sense, it tends to prove what we already suspected, that the GN$93$ abundances should not be used in the spectroscopic determination of the $\left(\frac{Z}{X}\right)$ for this study. This emphasizes the importance of consistency with the differential spectroscopic study that determined the [Fe/H] value for the star one wishes to study. In this particular case, we see that the inversion of $t_{u}$ is able to detect such inconsistencies, thanks to its sensitivity on metallicity mismatches. However, if the model is build with the $Z/X$ determined from the AGSS$09$ solar reference value, but using the GN$93$ solar heavy element mixture, we cannot detect inconsistencies. In fact, we obtain the same conclusion as before since these models are nearly identical in terms of internal structure.
\subsection{Impact on the mass and radius dispersion of $16$CygB}
In the previous section, we used the $t_{u}$ inversion to reduce the age, mass and radius dispersion of $16$CygA. Moreover, we know from Sect. \ref{secrescygb} that the inversion of $t_{u}$ for $16$CygB can only be used to check the consistency of the model but not to gain additional information. However, since these stars are binaries, we can say that the age values of the models $16$CygB must be compatible with those obtained for $16$CygA. From the inversion results of $16$CygA, we have also deduced that we had to include atomic diffusion in the stellar models and since both stars are very much alike, there is no reason to discard microscopic diffusion from the models of the $B$ component when we know that it has to be included in the models for the $A$ component.
\\
\\
Therefore, we can ask the question of what would the mass and radius of $16$CygB be if one includes diffusion as in $16$CygA and ensures that the age of the models remain compatible. The question of the chemical composition is also important since \citet{Ramirez} found a somewhat lower value for the [Fe/H] of the $B$ component and \citet{Verma} found larger uncertainties for the surface helium abundance, although the centroid value was the same as that of $16$CygA.
\begin{table}[t]
\centering
\caption{Accepted parameters obtained for $16$CygB when taking into account the constraints on $16$CygA.}
\label{tabcygBfinalfit}
\begin{tabular}{r | c}
\hline \hline
& \textbf{Accepted $16$CygB models} \\
\hline
$M$ $(\mathrm{M_{\odot}})$& $0.93-0.96$ \\
Age $(\mathrm{Gy})$& $7.0-7.4$\\
$Y_{0}$ $\mathrm{(dex)}$& $0.30-0.31$\\
$Z_{0}$ $\mathrm{(dex)}$& $0.0151-0.0186$\\
$D$ & $1.00-1.15$ \\
$\alpha_{\mathrm{MLT}}$ & $1.65-1.80$ \\
$L$ $(\mathrm{L_{\odot}})$ & $1.17-1.24$ \\
$R$ $(\mathrm{R_{\odot}})$& $1.08-1.10$ \\
\hline
\end{tabular}
\end{table}
To build these new models, we imposed that they included atomic diffusion with a coefficient $D$ of $1.0$ or $1.15$. The age was to be between $7.0$ $Gy$ and $7.4$ $Gy$. The metallicity was imposed to be within the error bars provided by \citet{Ramirez} and the surface helium abundance was to be within $\left[0.24, 0.25 \right]$. We used the same constraints as before to carry out the fits using the Levenberg-Marquardt algorithm and found that the mass was to be within $0.93$ $M_{\odot}$ and $0.96$ $M_{\odot}$, thus a $1.5 \%$ dispersion and the radius was to be within $1.08$ $R_{\odot}$ and $1.10$ $R_{\odot}$, thus a $1 \%$ dispersion. We would like to emphasize here that these values are of course dependent on the results of the modelling of $16$CygA and are thus more model-dependent since they do not result from constraints obtained through seismic inversions but are a consequence of the binarity of the system. It is clear that a changes in the values of the fundamental parameters for $16$CygA will induce a change in the values of $16$CygB.

\subsection{Discussion}
The starting point of this study was the determination of fundamental parameters for both $16$CygA and $16$CygB using seismic, spectroscopic and interferometric constraints. However, the differences between our results and those from \citet{Metcalfe} raise questions. One could argue that the inversion leads to problematic results and that the diagnostic would have been different if the surface helium determination from \citet{Verma} would have not been available. 
\\
\\
Therefore, for the sake of comparison, we asked the question of what would have been the results of this study if we had not included the surface helium abundance from \citet{Verma} in the model selection process. We carried out a few supplementary fits, using the mass, age, $\alpha_{MLT}$, $X_{0}$ and $Z_{0}$ as free parameters, using all the previous observational constraints as well as the prescription for microscopic diffusion from \citet{Thoul}, but excluding the $Y_{f}$ value. The results speak for themselves since we end up with a model having a mass of $1.09M_{\odot}$ and an age of $7.19Gy$ compatible with the results from \citet{Metcalfe}. This means that the determining property that leads to the changes in the fundamental parameters of the star was, as previously guessed, the surface helium value. Hence, without this $Y_{f}$ constraint, one would end up with two solutions with completely different masses and ages but fitting the same observational constraints. This does not mean that the results from \citet{Metcalfe} are wrong, they simply were the best results one could obtain without the surface helium constraint and with $3$ months of Kepler data. In fact, this is only an illustration of the importance of chemical composition constraints in stellar physics. The $Y_{0}-M$ trend has already been described in \citet{Baudin} and the fact that we find lower masses when increasing the helium abundance is, ultimately, no surprise.
\\
\\
At this point, we wanted to know what the inversion results would have been if we had used reference models with similar parameters as what was obtained in \citet{Metcalfe}. We ended up with similar results for both the acoustic radius and the mean density inversion, but more interestingly, the $t_{u}$ inversion also provided non-negligible corrections for this model. In fact, even with microscopic diffusion, the $\frac{t_{u,ref}}{R_{Ref}^{6}}$ value was of: $2.72 g/cm^{6}$ whereas the inverted result was: $\frac{t_{u,inv}}{R_{obs}^{6}}=3.5 \pm 0.5 g/cm^{6}$. Therefore the diagnostic potential of the indicator is still clear, since it could have provided indications for a change in the core structure of the model. Assuming that diffusion velocities are around $20 \%$ accurate, one could have invoked either an extra-mixing process or a change in the initial helium composition to explain this result. Disentangling between both cases would then have probably required additional indicators.
\section{Conclusion}\label{secconclusion}
In this article, we have applied the inversion techniques presented in a series of previous papers to the binary system $16$CygA and $16$CygB. The first part of this study consisted in determining suitable reference models for our inversion techniques. This was done using a Levenberg-Marquardt algorithm and all the seismic, spectroscopic and interferometric observational constraints available. We used the oscillation frequencies from \citet{Davies}, the interferometric radii from \citet{White}, the spectroscopic constraints from \citet{Ramirez} and the surface helium constraints from \citet{Verma}. 
\\
\\
Due to these constraints on the surface chemical composition, our results are different from those of \citet{Metcalfe}. The test case we made without using the constraint on surface helium from \citet{Verma} demonstrates the importance of constraints on the chemical composition for seismic studies. In fact, having to change the initial helium abundance from $0.25$ to values around $0.3$ is of course not negligible. This emphasizes that we have to be careful when using free parameters for the stellar chemical composition in seismic modelling. The same can be said for the constraints on the stellar [Fe/H] from the study of \citet{Ramirez}. For this particular constraint, we have to add the importance of the solar mixture used in the spectroscopic study. Due to the important changes in the $\left(\frac{Z}{X} \right)_{\odot}$ from the GN$93$ abundances to the AGSS$09$ abundances, consistency with the spectroscopic study has to be ensured. Otherwise, the final results of the seismic modelling could be biased. In this case, we chose to use the AGSS$09$ solar mixture. We note that our reference models tend to be consistent with the spectroscopic, seismic and interferometric constraints and that independent modelling of both stars leads to consistent ages. We also note the presence of a certain modelling degeneracy in terms of chemical composition and microscopic diffusion. Accordingly, we could obtain rather different values for the mass, the radius and the age of both stars by assuming more intense diffusion and changing the chemical composition within the error bars from both \citet{Ramirez} and \citet{Verma}. We also note that when not considering the constraints on surface helium, we obtained results compatible with \citet{Metcalfe} but the $t_{u}$ values were too low even when diffusion was included in the models. This reinforces the importance of constraints on the chemical composition and illustrates to what extent inversions could be used given their intrinsic limitations.
\\
\\
Having obtained suitable reference models, we then carried out inversions for the mean density , $\bar{\rho}$, the acoustic radius, $\tau$, and a core condition indicator, $t_{u}$. The first two quantities were used to improve the quality of the reference models. As a by product, we noted that models fitting both $\bar{\rho}$ and $\tau$ were in better agreement in terms of individual frequencies. We also found that both of these quantities could not disentangle the effect of the degeneracy in terms of diffusion and chemical composition. However, they could be well suited to analyse uppers layers along with other quantities.
\\
\\
After the second modelling process, we carried out inversion for the $t_{u}$ indicator and noted that the degeneracy in terms of chemical composition and diffusion could be reduced for $16$CygA. In fact, to agree with the inverted result, one has to consider diffusion speed calibrated for the sun or slightly higher (by $10 \%$ or $15 \%$). Values higher than $20 \%$ were considered not physical by \citet{Thoul} and were therefore not analysed in this study. At the end of day, we come up with a smaller dispersion in terms of mass and age for $16$CygA, namely that this component should have a mass between $0.97M_{\odot}$ and $1.0M_{\odot}$, a radius between $1.188R_{\odot}$ and $1.200R_{\odot}$ and an age between $7.0Gy$ and $7.4Gy$. Again the slight differences between the seismic radius provided here and the interferometric radius might stem from different definitions of the interferometric radius and the seismic one. We also conclude that the $t_{u}$ inversion for $16$CygB could only be used as a consistency check but could not help reduce the dispersion in age. However, as these stars are binaries, a reduced age dispersion for one component means that the second has to be consistent with this smaller dispersion. Therefore, we were able to deduce a smaller mass and radius dispersion for the second component, namely between $0.93$ $M_{\odot}$ and $0.96$ $M_{\odot}$ and between $1.08$ $R_{\odot}$ and $1.10$ $R_{\odot}$.
\\
\\
Finally, we draw the attention of the reader to the following points: the age dispersion we obtain is not model-independent, we assumed physical properties for the models and assumed that the agreement in $t_{u}$ was to be improved by varying the chemical composition within the observational constraints and by calibrating microscopic diffusion. This does not mean that no other mixing process has taken place during the evolutionary sequence that could somehow slightly bias our age determination. In that sense, further improved studies will be carried on, using additional structural quantities, more efficient global minimization tools for the selection of the reference models and possibly improved physical ingredients for the models. In conclusion, we show in this study that inversions are indeed capable of improving our use of seismic information and therefore through synergies with stellar modellers, helping us build new generations of more physically accurate stellar models.
\begin{acknowledgements}
G.B. is supported by the FNRS (``Fonds National de la Recherche Scientifique'') through a FRIA (``Fonds pour la Formation à la Recherche dans l'Industrie et l'Agriculture'') doctoral fellowship. D.R.R. is currently funded by the European Community's Seventh Framework Programme (FP7/2007-2013) under grant
agreement no. 312844 (SPACEINN), which is gratefully acknowledged. This article made use of an adapted version of InversionKit, a software developed in the context of the HELAS and SPACEINN networks, funded by the European Commissions's Sixth and Seventh Framework Programmes. We would also like to thank Guy Davies for his advice. 
\end{acknowledgements}
\bibliography{biblioarticle3}
\appendix
\section{Intermediate results of the forward modelling process}
After the first step of forward modelling, we carried out supplementary fits to obtain new reference models for both $16$CygA and $16$CygB. In fact, we replaced the average large frequency separation by the acoustic radius and the mean density, as discussed in Sect. \ref{secNewRef}. We used the following naming convention for these models: the first letter, $A$ or $B$ is associated with the star, namely $16CygA$ or $16CygB$; the second letter is associated with the chemical composition box in the right pannel of Fig. \ref{resultsbox}, $C$ being the central chemical composition, $L$ the left-hand side, $R$ the right-hand side, $U$ the upper side and $D$ the lower side ($D$ for down); the number $1$ or $2$ is associated with diffusion, $1$ is for models without microscopic diffusion, $2$ is for models including the prescriptions of \citet{Thoul} for microscopic diffusion. These results are illustrated in the following tables for both stars: 
\begin{table*}[t]
\caption{Optimal parameters obtained for $16$CygA using the acoustic radius and the mean density rather than $< \Delta \nu >$.}
\label{tabsecfitA}
  \centering
\begin{tabular}{r | c | c | c |  c  | c | c | c | c | c | c }
\hline \hline
 & \textbf{S$_{A,C1}$}& \textbf{S$_{A,C2}$}  
 & \textbf{S$_{A,U1}$}& \textbf{S$_{A,U2}$} 
 & \textbf{S$_{A,D1}$}& \textbf{S$_{A,D2}$} 
 &\textbf{S$_{A,R1}$}& \textbf{S$_{A,R2}$}
 & \textbf{S$_{A,L1}$}& \textbf{S$_{A,L2}$} \\ \hline
 \textit{M ($\mathrm{M_{\odot}}$)}& $1.049$ & $0.999$ 
 & $1.039$ & $0.994$&$1.06$&$1.007$&$1.055$&$1.001$&$1.049$&$0.983$\\
\textit{R ($\mathrm{R_{\odot}}$)}& $1.221$ & $1.201$ 
&$1.216$ &$1.198$&$1.227$&$1.203$&$1.222$&$1.201$&$1.220$&$1.195$\\ 
\textit{Age ($\mathrm{Gyr}$)} &$8.30$& $7.38$ 
&$8.09$ & $6.77$&$8.33$&$7.53$&$8.34$&$7.31$&$8.11$&$7.33$\\ 
\textit{$T_{\mathrm{eff}}$ ($\mathrm{K}$)} & $5852$ & $5828$ 
& $5903$ & $5992$&$5842$&$5811$&$5827$&$5837$&$5912$&$5877$\\ 
\textit{$L_{\odot}$ $(\mathrm{L_{\odot}})$} & $1.570$ & $1.494$
& $1.61255$ & $1.66196$&$1.574$&$1.482$&$1.546$&$1.504$&$1.633$&$1.529$\\
\textit{$Z_{0}$} & $0.0165$& $0.0205$ 
& $0.0162$ & $0.0195$& $0.0167$&$0.020$&$0.0174$&$0.0210$&$0.0155$&$0.0188$\\
\textit{$Y_{0}$} &$0.24$& $0.2945$ 
&$0.25$ & $0.307729$& $0.23$&$0.28609$&$0.24$&$0.2968$&$0.24$&$0,299$\\
\textit{$\alpha_{\mathrm{MLT}}$} & $1.68$ & $1.74$ 
& $1.75$ &$1.97$&$1.68957$&$1.72$&$1.67$&$1.76$&$1.75$&$1.78$\\
\textit{$D$} & $0.0$& $1.0$ 
&$0.0$ & $1.0$& $0.0$ &$1.0$&$0.0$&$1.0$&$0.0$&$1.0$\\
\hline
\end{tabular}
\end{table*}

\begin{table}[t]
\caption{Optimal parameters obtained for $16$CygB using the acoustic radius and the mean density rather than $< \Delta \nu >$.}
\label{tabsecfitB}
  \centering
\begin{tabular}{r | c | c }
\hline \hline
 & \textbf{S$_{B,C1}$}& \textbf{S$_{B,C2}$} \\ \hline
 \textit{M ($\mathrm{M_{\odot}}$)}& $1.00793$ &$0.960578$ \\
\textit{R ($\mathrm{R_{\odot}}$)}& $1.10560$ & $1.08821$ \\ 
\textit{Age ($\mathrm{Gyr}$)} &$8.16170$&$7.23565$  \\ 
\textit{$T_{\mathrm{eff}}$ ($\mathrm{K}$)} & $5793$ & $5829$ \\ 
\textit{$L_{\odot}$ $(\mathrm{L_{\odot}})$} & $1.23498$ & $1.22672$ \\
\textit{$Z_{0}$} & $0.0151$&$0.01805$\\
\textit{$Y_{0}$} &$0.24$&$0.29183$  \\
\textit{$\alpha_{\mathrm{MLT}}$} & $1.667$ & $1.78034$ \\
\textit{$D$} & $0.0$ &$1.0$ \\
\hline
\end{tabular}
\end{table}
\end{document}